# Neutral molecular cluster formation of sulfuric acid–dimethylamine observed in real time under atmospheric conditions

Andreas Kürten[a,1], Tuija Jokinen[b], Mario Simon[a], Mikko Sipilä[b,c], Nina Sarnela[b], Heikki Junninen[b], Alexey Adamov[b], João Almeida[a,d], Antonio Amorim[e], Federico Bianchi[f], Martin Breitenlechner[g,h], Josef Dommen[f], Neil M. Donahue[i], Jonathan Duplissy[b,c,d], Sebastian Ehrhart[a,d], Richard C. Flagan[j], Alessandro Franchin[b], Jani Hakala[b], Armin Hansel[g,h], Martin Heinritzi[a,h], Manuel Hutterli[k], Juha Kangasluoma[b], Jasper Kirkby[a,d], Ari Laaksonen[l,m], Katrianne Lehtipalo[b,n], Markus Leiminger[a], Vladimir Makhmutov[o], Serge Mathot[d], Antti Onnela[d], Tuukka Petäjä[b], Arnaud P. Praplan[b], Francesco Riccobono[f], Matti P. Rissanen[b], Linda Rondo[a], Siegfried Schobesberger[b], John H. Seinfeld[j], Gerhard Steiner[b,h], António Tomé[e], Jasmin Tröstl[f], Paul M. Winkler[p], Christina Williamson[a], Daniela Wimmer[a,b], Penglin Ye[i], Urs Baltensperger[f], Kenneth S. Carslaw[q], Markku Kulmala[b,c], Douglas R. Worsnop[b,r], and Joachim Curtius[a]

[a]Institute for Atmospheric and Environmental Sciences, Goethe-University of Frankfurt, 60438 Frankfurt am Main, Germany; [b]Department of Physics, University of Helsinki, 00014 Helsinki, Finland; [c]Helsinki Institute of Physics, University of Helsinki, 00014 Helsinki, Finland; [d]European Organization for Nuclear Research, CH-1211 Geneva, Switzerland; [e]Laboratory for Systems, Instrumentation and Modeling for Space and the Environment, University of Lisbon and University of Beira Interior, 1749-016 Lisbon, Portugal; [f]Laboratory of Atmospheric Chemistry, Paul Scherrer Institute, 5232 Villigen, Switzerland; [g]Ionicon Analytik GmbH, 6020 Innsbruck, Austria; [h]Institute for Ion Physics and Applied Physics, University of Innsbruck, 6020 Innsbruck, Austria; [i]Center for Atmospheric Particle Studies, Carnegie Mellon University, Pittsburgh, PA 15213; [j]Division of Chemistry and Chemical Engineering, California Institute of Technology, Pasadena, CA 91125; [k]Tofwerk AG, 3600 Thun, Switzerland; [l]Department of Applied Physics, University of Eastern Finland, 70211 Kuopio, Finland; [m]Finnish Meteorological Institute, 00101 Helsinki, Finland; [n]Airmodus Ltd., 00560 Helsinki, Finland; [o]Solar and Cosmic Ray Research Laboratory, Lebedev Physical Institute, 119991 Moscow, Russia; [p]Faculty of Physics, University of Vienna, 1090 Vienna, Austria; [q]School of Earth and Environment, University of Leeds, Leeds LS2 9JT, United Kingdom; and [r]Aerodyne Research Inc., Billerica, MA 01821

**For atmospheric sulfuric acid (SA) concentrations the presence of dimethylamine (DMA) at mixing ratios of several parts per trillion by volume can explain observed boundary layer new particle formation rates. However, the concentration and molecular composition of the neutral (uncharged) clusters have not been reported so far due to the lack of suitable instrumentation. Here we report on experiments from the Cosmics Leaving Outdoor Droplets chamber at the European Organization for Nuclear Research revealing the formation of neutral particles containing up to 14 SA and 16 DMA molecules, corresponding to a mobility diameter of about 2 nm, under atmospherically relevant conditions. These measurements bridge the gap between the molecular and particle perspectives of nucleation, revealing the fundamental processes involved in particle formation and growth. The neutral clusters are found to form at or close to the kinetic limit where particle formation is limited only by the collision rate of SA molecules. Even though the neutral particles are stable against evaporation from the SA dimer onward, the formation rates of particles at 1.7-nm size, which contain about 10 SA molecules, are up to 4 orders of magnitude smaller compared with those of the dimer due to coagulation and wall loss of particles before they reach 1.7 nm in diameter. This demonstrates that neither the atmospheric particle formation rate nor its dependence on SA can simply be interpreted in terms of cluster evaporation or the molecular composition of a critical nucleus.**

## Significance

A significant fraction of atmospheric aerosols is formed from the condensation of low-volatility vapors. These newly formed particles can grow, become seeds for cloud particles, and influence climate. New particle formation in the planetary boundary layer generally proceeds via the neutral channel. However, unambiguous identification of neutral nucleating clusters has so far not been possible under atmospherically relevant conditions. We explored the system of sulfuric acid, water, and dimethylamine in a well-controlled laboratory experiment and measured the time-resolved concentrations of neutral clusters. Clusters containing up to 14 sulfuric acid and 16 dimethylamine molecules were observed. Our results demonstrate that a cluster containing as few as two sulfuric acid and one or two dimethylamine molecules is already stable against evaporation.

Aerosol particles are important constituents of the Earth's atmosphere. A large fraction of the particles form by nucleation of low-volatility vapors. The newly formed particles first consist only of a few molecules and have diameters between 1 and 2 nm. Under favorable conditions, where loss rates are small and growth rates are sufficiently large, the particles can reach sizes of ∼50 nm where they can act as cloud condensation nuclei (CCN). Model simulations suggest that globally about half the CCN originate from new particle formation (NPF) (1). Therefore, NPF is an important process affecting the climate through the cloud albedo effect (2). Numerous studies have shown that sulfuric acid (SA; $H_2SO_4$) and water vapor are important compounds participating in the formation of new particles (3, 4). However, atmospheric boundary layer NPF events cannot be explained by the binary nucleation of these two compounds alone (5); therefore, at least one additional substance besides SA and water vapor is required (6). However, the chemical identity of the compounds responsible for the high observed NPF rates remains to be fully elucidated. Recent theoretical (7, 8) and experimental (6, 9–15) studies have shown that amines efficiently stabilize SA

Author contributions: A.K., M. Simon, M. Sipilä, A. Adamov, F.B., J. Dommen, N.M.D., J. Duplissy, S.E., R.C.F., A.F., J.H., A.H., J. Kirkby, A.L., V.M., S.M., A.O., T.P., A.P.P., F.R., M.P.R., S.S., J.H.S., G.S., A.T., D.W., U.B., K.S.C., M.K., D.R.W., and J.C. designed research; A.K., T.J., M. Simon, M. Sipilä, N.S., A. Adamov, J.A., A. Amorim, F.B., M.B., J. Duplissy, S.E., J.H., M. Heinritzi, J. Kangasluoma, J. Kirkby, K.L., M.L., A.P.P., F.R., M.P.R., L.R., S.S., G.S., A.T., J.T., P.M.W., C.W., D.W., and P.Y. performed research; A.K., T.J., M. Simon, M. Sipilä, H.J., M.B., A.H., and M. Hutterli contributed new reagents/analytic tools; A.K., T.J., M. Simon, M. Sipilä, N.S., H.J., J.A., F.B., M.B., J. Dommen, M. Heinritzi, J. Kirkby, K.L., A.P.P., S.S., G.S., A.T., and D.R.W. analyzed data; and A.K. and J.C. wrote the paper.

The authors declare no conflict of interest.

*This Direct Submission article had a prearranged editor.

Freely available online through the PNAS open access option.

[1]To whom correspondence should be addressed. Email: kuerten@iau.uni-frankfurt.de.

clusters and can also form new particles together with methanesulfonic acid and water vapor (16). Other research involving amines focused on the physiochemical properties of alkylaminium sulfates (17) and on the substitution of ammonia by amines in sulfuric acid–ammonia clusters (18, 19). A recent review article summarizes the findings on the atmospheric implications of amines (20). Oxidized organic compounds can also contribute to the enhancement of NPF rates (21–25). Because neutral (uncharged) new particle formation appears to dominate in the boundary layer (26, 27), it is critically important to develop techniques to measure the composition of small neutral clusters as they grow from monomers to ultrafine particles. New particle formation is highly nonlinear with respect to the concentration of the precursor gases; therefore, it is essential for these measurements to be conducted at the (extremely low) concentrations of relevant precursor gases found in the atmosphere.

Considerable progress has been made in recent years toward the development of instruments for measurement of gaseous compounds and particles during NPF events. The number density of small particles down to ~1.2 nm in diameter can now be measured with the particle size magnifier (28). For charged clusters, the molecules involved can be measured with the Atmospheric Pressure interface–Time Of Flight (APi-TOF) mass spectrometer (29). However, an instrument to measure the precise molecular composition of neutral clusters has been developed only very recently (30). Although atmospheric neutral clusters have previously been detected (12, 13, 31), their molecular composition was not unambiguously resolved.

Here we present results using a Chemical Ionization (CI)–APi-TOF mass spectrometer that can resolve the elemental composition of neutral clusters up to ~2,000 atomic mass units (30). The largest detected clusters have a mobility diameter around 2 nm, which falls within the measurement range of recently developed condensation particle counters (28). Thus, the CI-APi-TOF can measure the molecular composition of neutral clusters from the molecular up to the macroscopic size. The results shown here relate to a previous study conducted at the Cosmics Leaving Outdoor Droplets (CLOUD) chamber at the European Organization for Nuclear Research (CERN), which found that atmospheric boundary layer nucleation rates have the same range of values as particle formation rates from sulfuric acid, dimethylamine, and water (15). Almeida et al. (15) reported particle formation rates measured by condensation particle counters and showed results for charged clusters measured with an APi-TOF mass spectrometer (29). One of the conclusions was, however, that ions are not essential in the formation of particles in the SA–dimethylamine (DMA) system when nucleation rates exceed a certain value. Our study reports on the very first measurements to our knowledge of neutral clusters made with a CI-APi-TOF.

## Results and Discussion

Ternary nucleation of SA, DMA, and water was studied using the CLOUD chamber at atmospherically relevant concentrations of SA, between ~$5 \times 10^5$ and $1.5 \times 10^7$ $cm^{-3}$, and with DMA [$(CH_3)_2NH$] mixing ratios between 5 and 32 parts per trillion by volume (pptv). The experiments were conducted at 278 K and 38% relative humidity. Two CI-APi-TOF mass spectrometers were deployed, both using nitrate charger ions (*SI Text* and Fig. S1).

Recent results obtained at CLOUD for the SA–DMA system showed that charged clusters containing three or more SA molecules (including the core bisulfate ion) grow by maintaining a near 1:1 ratio between the SA and DMA molecules during ion-induced nucleation (15). Here we present the first measurements, to our knowledge, of the neutral nucleating clusters. For these experiments a high-voltage clearing field is applied inside the CLOUD chamber to remove all ions and charged clusters that form due to galactic cosmic rays (5). The neutral clusters are sampled from the chamber; they only become charged upon entering the nitrate charging unit of the CI-APi-TOF instrument. The elemental composition of clusters is unambiguously identified from their exact mass-to-charge ratio due to the high mass accuracy (typically better than 10 ppm) and resolving power (maximum of 4,500 Th/Th) of the instrument. Isotopic ratios provide additional information to help resolve between atomic species.

The neutral clusters are seen to grow by stepwise addition of one SA followed by one DMA molecule (Fig. 1 *A* and *B*), according to the same base stabilization mechanism seen previously for charged clusters (15). The largest detected neutral cluster contains 14 SA and 16 DMA molecules. Some of the smallest clusters [tetramer and smaller, i.e., $HSO_4^-(H_2SO_4)_k$ where $k \leq 3$] are measured without DMA. However, it is likely that at least one DMA molecule is lost from the neutral clusters during the charging process because both $HSO_4^-$ and $NO_3^-$ are Lewis bases which compete with DMA to attach to a SA molecule. On the other hand, if not all $HNO_3$ evaporates from the cluster in the charging process, the evaporation of DMA can be prevented. This is indeed observed for the SA dimers [$HSO_4^-(H_2SO_4)$], which are detected with up to two DMA molecules (Fig. 1*B*). This is the first time to our knowledge that a stabilizing compound has been directly observed in the SA dimer. Additional evidence for the stabilizing effect of DMA on the dimers is provided by the magnitude of the dimer signal. For a binary system (SA and water), the dimer concentration is expected to be at least six orders of magnitude lower than seen in our measurements (32). Therefore, only stabilization by DMA can explain such high dimer concentrations (15, 33). Finally, we add that quantum chemical calculations suggest that even the neutral monomer of SA is bound to a DMA molecule (8), although this cannot yet be confirmed experimentally because DMA rapidly evaporates from the bisulfate ion.

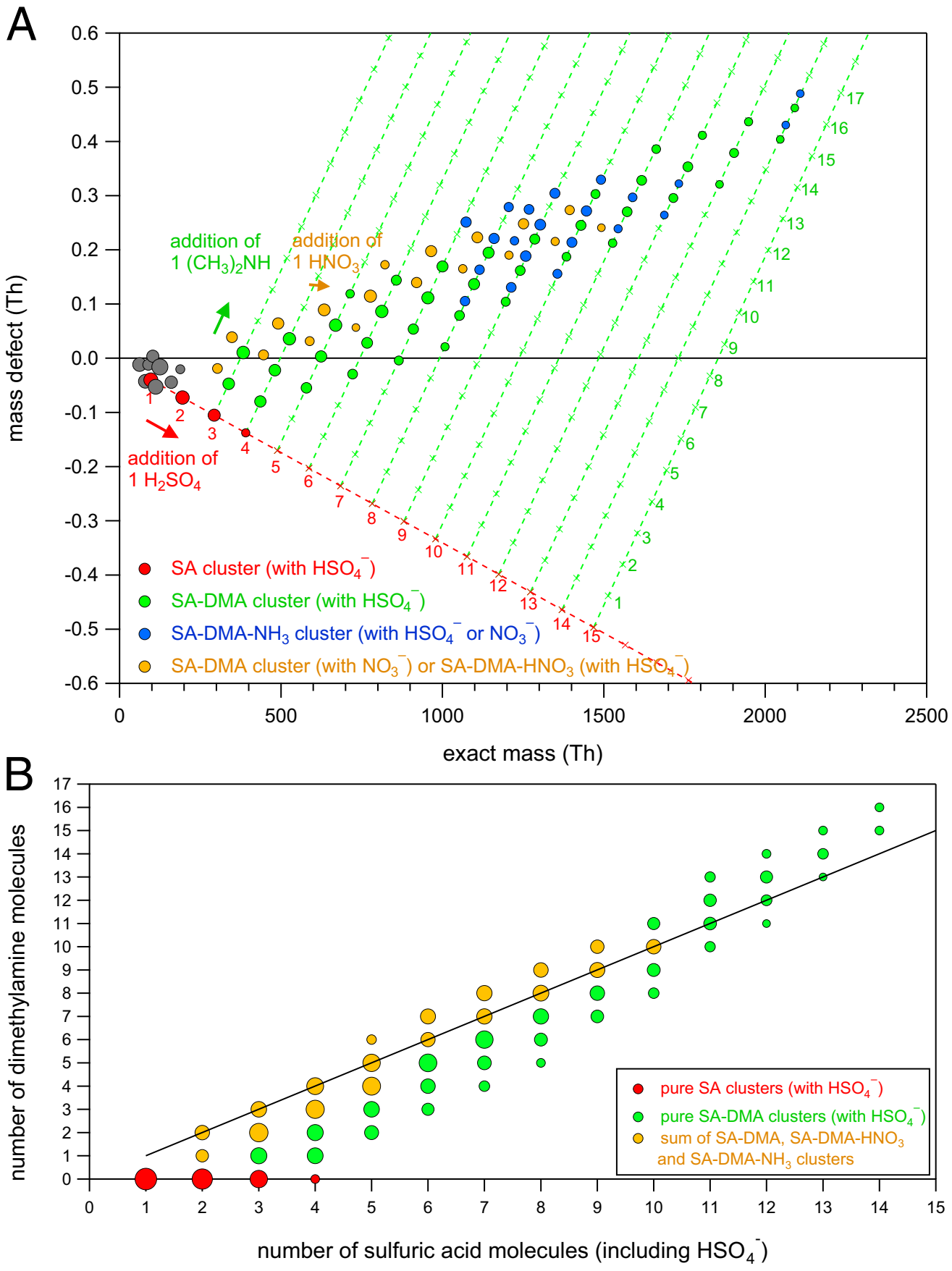

**Fig. 1.** Molecular weight and composition of neutral clusters during a new particle formation event with SA and DMA. (*A*) Mass defect plot measured with the CI-APi-TOF mass spectrometer for an experiment with 10 pptv DMA and $3 \times 10^7$ $cm^{-3}$ SA. The mass defect is the difference between the exact mass and the nominal (integer) mass of the cluster. The symbol size is proportional to the logarithm of the signal intensity (count rate). Background ions and clusters (not containing SA or DMA) are indicated by gray symbols. (*B*) The same data with the signals grouped to show the number of SA and DMA molecules in the cluster.

The temporal evolution of the cluster concentrations in a single representative experiment is shown in Fig. 2. The experiment is started by turning on the UV illumination; this initiates SA production and leads to sequential appearance of progressively larger clusters. Each cluster reaches a steady-state concentration when its production and loss rates are equal. The cluster concentrations predicted by a kinetic model with unit sticking efficiency are also shown in Fig. 2. The only free parameter in the model is the monomer production rate, which was adjusted to fit the measured monomer concentration ($N_1$). The modeled dimer concentration ($N_2$) matches the experimental value within a factor of 1.5, which is within the uncertainties of the detection efficiency. The uncertainty in the measured trimer ($N_3$), tetramer ($N_4$), and pentamer ($N_5$) concentrations increases progressively (*SI Text* and Fig. S2), and the modeled concentrations of these clusters are systematically higher than the measured values.

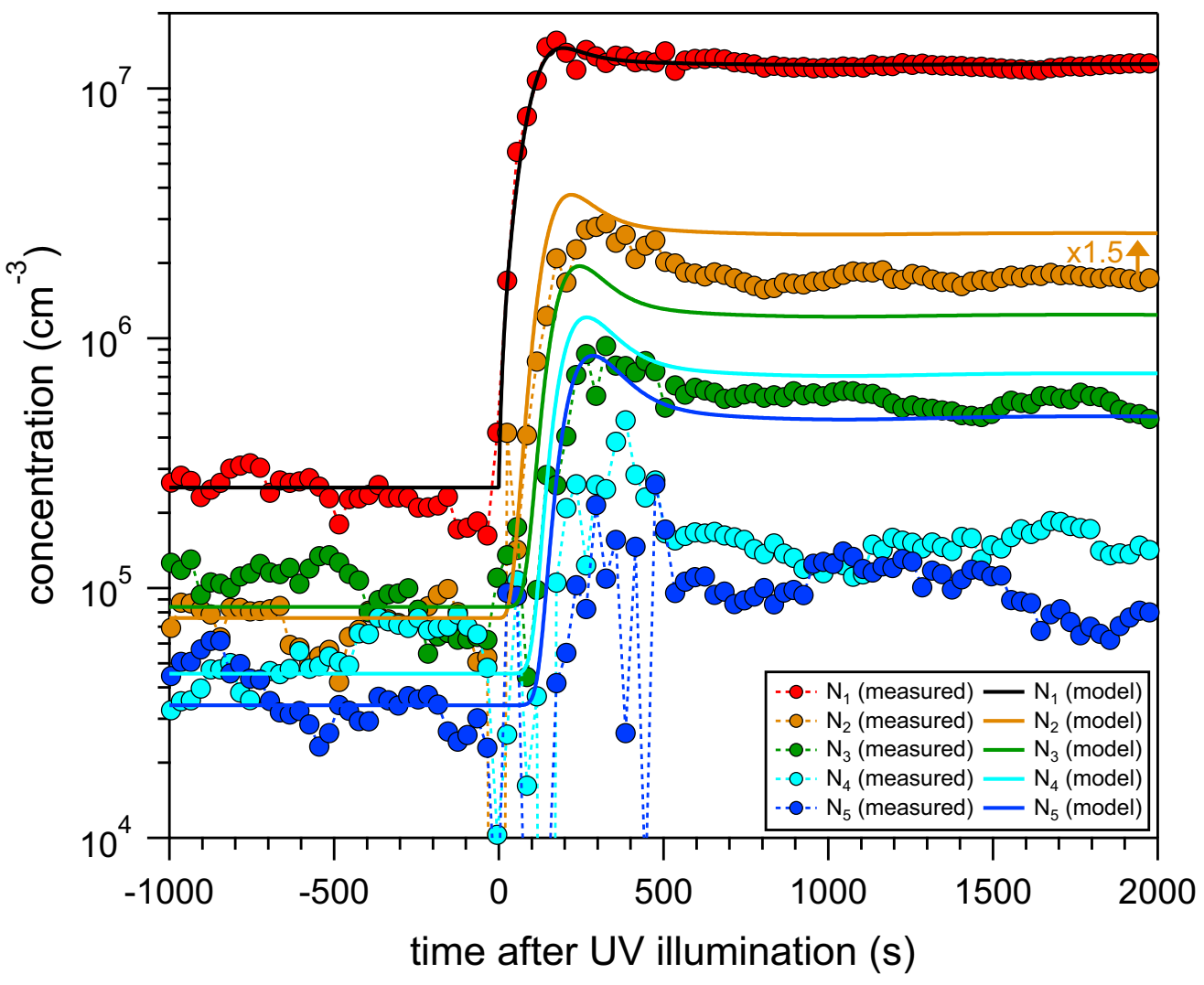

**Fig. 2.** Time evolution of measured and modeled neutral cluster concentrations. The concentrations were measured (circles and dashed lines) with the CI-APi-TOF mass spectrometer during an experiment with 20 pptv DMA and $1.2 \times 10^7$ cm$^{-3}$ SA. The notations $N_1$ (monomer), $N_2$ (dimer), $N_3$ (trimer), $N_4$ (tetramer), and $N_5$ (pentamer) refer to the number of sulfuric acid molecules in the cluster. To obtain the cluster concentrations ($N_2$ to $N_5$), the signals with different amounts of DMA but with a certain number of sulfur atoms are added up (*SI Text*). The experiment is started at zero time by illuminating the chamber with UV light, which initiates production of sulfuric acid monomers ($N_1$). The measurements are compared with calculated cluster concentrations (solid lines) from a kinetic model, which assumes that cluster evaporation rates are zero and that each collision leads to cluster growth (unit sticking efficiency). The discrepancy between the modeled and measured steady-state $N_2$ is indicated as a factor of ~1.5. The increasing offset between measured and modeled data with increasing cluster size is consistent with a declining transmission efficiency in the CI-APi-TOF. The excellent agreement on the timing of each successive cluster indicates negligible SA evaporation, confirming that the cluster growth is kinetically limited. The time resolution is 30 s. The data are smoothed by a running average except for the period where the cluster concentrations rise strongly (between 0 and 500 s).

However, when normalizing the cluster concentrations to their steady-state values, comparison between measured and modeled concentrations assuming zero evaporation shows excellent agreement (Figs. S3 and S4). Therefore, these data indicate that NPF in the SA–DMA system is very likely kinetically controlled; that is, cluster evaporation rates are effectively zero. In the present work, the term "kinetically limited nucleation" refers to a situation where the cluster growth is entirely controlled by collisions with monomers and smaller clusters and not by the evaporation of monomers. Because the measured $N_2$ agrees with the kinetic limit calculation, this should apply also for the larger clusters because evaporation rates are expected to decrease with increasing cluster size (13, 15). The discrepancies between the modeled and measured $N_k$ (for $k \geq 3$) in Fig. 2 can be explained by the uncertainties in deriving the concentrations from the measured cluster signals, particularly because it is presently not possible to calibrate the CI-APi-TOF for cluster concentration measurements (*SI Text*). Therefore, the charging and detection efficiencies of the clusters are not well constrained and can account for the uncertainties.

The steady-state $N_2$, $N_3$, $N_4$, and $N_5$ concentrations are shown as a function of the SA monomer concentration ($N_1$) in Fig. 3. For comparison, the modeled cluster concentrations are shown for the kinetic limit and for finite dimer evaporation rates of 0.01 and 0.1 s$^{-1}$. The measurements agree closely with the modeled $N_2$ assuming no evaporation, providing further evidence that the dimer with DMA is stable. This contrasts with earlier Chemical Ionization Mass Spectrometer (CIMS) measurements in which $N_2$ was about a factor of 5 lower than expected for kinetic NPF (15). Fragmentation of the $HSO_4^-(H_2SO_4)$ ions in the CIMS is the most likely explanation for this discrepancy.

For the larger clusters the measured concentrations shown in Fig. 3 are below the kinetic limit calculations (zero evaporation rate). However, it is important to note that the slopes of the experimental cluster concentrations vs. $N_1$ are compatible with the kinetic limit curves but incompatible with those for finite evaporation rates. Evaporation of the clusters would generally lead to a steepening of the slopes, which is not observed. Furthermore, the model curves that go through the data require progressively larger values of the dimer evaporation rate ($k_{2,\mathrm{evap}}$), which is unphysical. In contrast, curves accounting for progressive decrease in the detection efficiency reproduce the shape of the observations with negligible SA evaporation (*SI Text*). Our measurements therefore indicate that the larger clusters also form at or near the kinetic limit. This confirms for the first time, to our knowledge, through direct measurement of neutral clusters that SA–DMA new particle formation can be a purely kinetic process. Kinetic behaviour contrasts with the observation of stable prenucleation clusters as precursors in crystallization (34). These clusters are stable up to a certain size before they reach a barrier where they become unstable before eventually crossing the nucleation barrier. Although we cannot rule out entirely that such a barrier could also exist in the SA–DMA system, it would almost certainly be beyond the pentamer. The reason is that otherwise, no monotonic decrease, but rather a sharp step, should become visible when plotting the ratio between measured and modeled cluster concentration as a function of cluster size (*SI Text* and Fig. S2). The data from Fig. 1*B* provide further qualitative information that even beyond the pentamer, no sharp drop in the cluster signal is visible for clusters as large as 2 nm in mobility diameter.

Chen et al. (13) concluded that nucleation of SA and base molecules (ammonia and/or amines) is not proceeding at the kinetic limit and that an evaporation rate of 0.4 s$^{-1}$ for the trimer containing at least one base molecule can explain the atmospheric particle formation rates in Atlanta and Mexico City. This contrasts with our finding that cluster evaporation rates are negligible in the SA–DMA system. However, one needs to keep in mind that the DMA concentration is also an important parameter. Indeed, the study by Chen et al. (13) showed for the SA–DMA system that the ratio between $N_2$ and $N_1$ reaches a constant value only when the DMA concentration is about a factor of 100 larger than the SA concentration. If this ratio is reached, the measured $N_2$ concentration approaches the expected value for the kinetic limit within errors (13). Additionally, the study by Almeida et al. (15) also shows this behavior where the nucleation rate is approaching a plateau when the DMA mixing ratio exceeds ∼10 pptv (i.e., ∼$2.5 \times 10^8$ $cm^{-3}$ at [$H_2SO_4$] = $2 \times 10^6$ $cm^{-3}$). One explanation for this plateauing effect is that SA•DMA agglomerates are required to initiate the formation of new particles (8, 15). However, quantum chemical calculations suggest that appreciable evaporation of SA•DMA is occurring (8). In this case, efficient dimer (and larger cluster) formation can only proceed if the arrival rate of DMA on a SA molecule is compatible with the evaporation rate of SA•DMA. Under these circumstances, collisions between SA and SA•DMA or between SA•DMA agglomerates can efficiently produce stable dimers and larger clusters. Therefore, the concentration of the stabilizing compound would determine what fraction of the clusters is stable, and the highest possible formation rates are only observed once the DMA to SA concentration ratio exceeds a certain threshold value. In this study the concentration ratio between DMA and SA was on average ∼110, and therefore, our results are consistent with previous observations where saturation effects at high DMA levels were observed (13, 15). However, reported atmospheric DMA mixing ratios are mostly below 10 pptv (35), therefore, it is unlikely that SA–DMA nucleation, if occurring, is always saturated with respect to DMA and can generally proceed at the kinetic limit. A similar process could occur in a system involving sulfuric acid and oxidized organics. However, it remains to be elucidated if these systems can produce equally stable neutral clusters with SA or other compounds. Another important aspect to be examined in the future is the effect of temperature, although quantum chemical calculations suggest that there is only a weak dependency for SA–DMA new particle formation on temperature (15). Using the same quantum chemical methods, this has also been concluded for the effect of relative humidity (RH). Varying the RH between 0 and 100% did only lead to a small increase of about a factor of 2 in the particle formation rates even when the DMA mixing ratio was as low as 0.1 pptv [at a SA concentration of $2 \times 10^6$ $cm^{-3}$ (15)]. Regarding the effect of water on the nucleation process, it is important to note that although SA and DMA molecules are not evaporating from the clusters, condensed water molecules can evaporate. However, our data indicate that this does not have a substantial effect on the SA–DMA cluster stability, and therefore, the nucleation process is termed "kinetically limited."

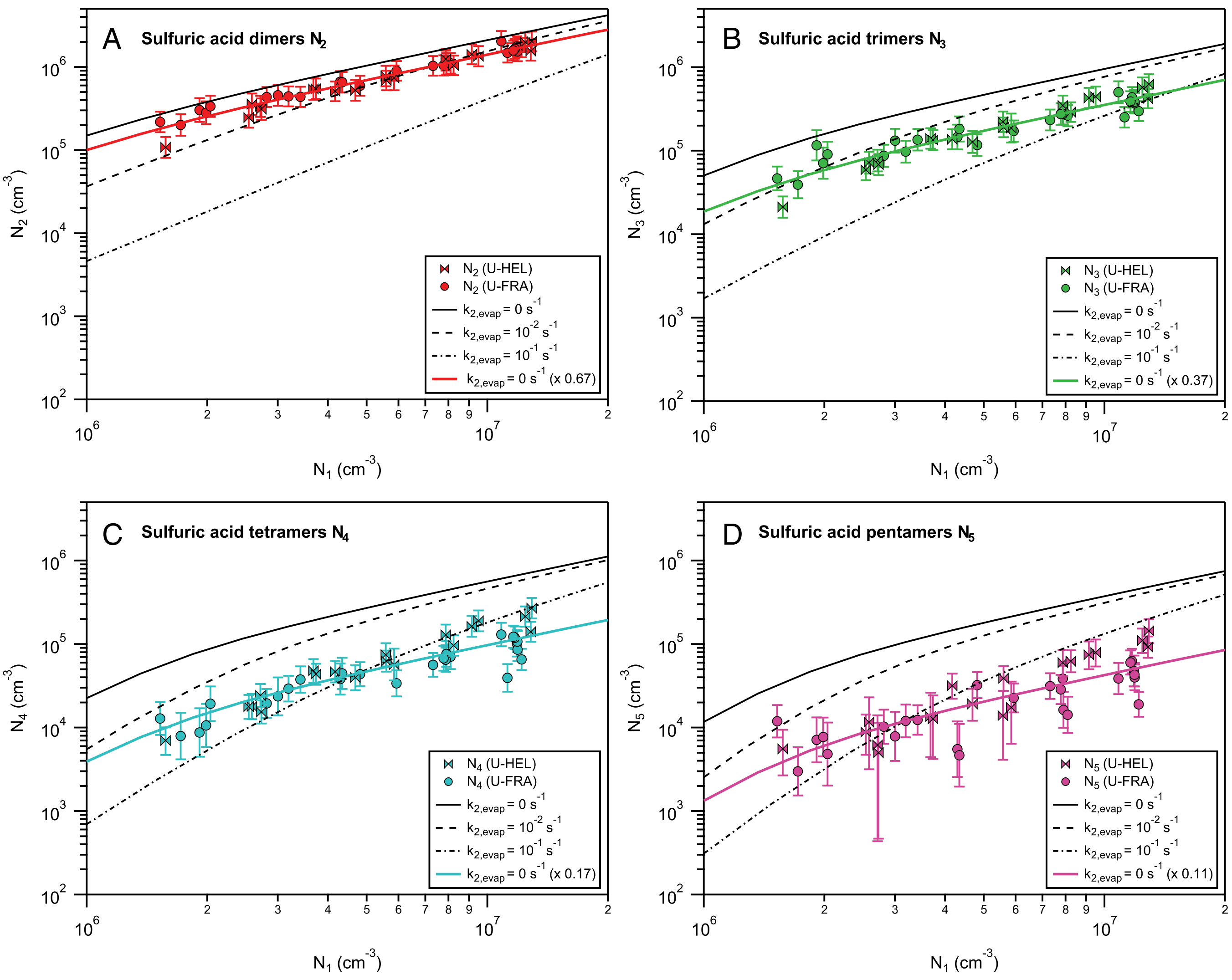

**Fig. 3.** Measured and modeled cluster concentrations [$N_2$ (*A*), $N_3$ (*B*), $N_4$ (*C*), and $N_5$ (*D*)] as a function of the sulfuric acid monomer concentration ($N_1$). The measured data are from two mass spectrometers (CI-APi-TOF-U-HEL and CI-APi-TOF-U-FRA; *SI Text*). Dimethylamine levels are between 5 and 32 pptv. Calculated steady-state cluster concentrations from a kinetic model are included. The solid lines assume that all evaporation rates are zero, whereas the dashed lines simulate a nonzero dimer evaporation rate ($k_{2,evap}$ = 0.01 or 0.1 $s^{-1}$). Error bars include the statistical variation plus a systematic error of ±30%. In addition, correction factors due to the unknown cluster charging and detection efficiencies need to be considered (Figs. S2 and S3). These are derived by scaling the model curve assuming no evaporation ($k_{2,evap}$ = 0 $s^{-1}$) to fit the measured data (colored lines). The derived scaling factors are 0.67 for $N_2$, 0.37 for $N_3$, 0.17 for $N_4$, and 0.11 for $N_5$ and are always in the direction of reducing the measurements below the true values. Using the model curve assuming zero dimer evaporation as the reference is justified because the slopes of the model curves with $k_{2,evap} > 0$ $s^{-1}$ do not match the slopes of the measured data. The uncertainty in $N_1$ (*x* axis) is a factor 1.5.

Molecular cluster measurements allow direct determination of the NPF rate at a given cluster size because at steady-state concentration, the formation (nucleation) rate equals the total loss rate (to larger clusters, chamber walls, etc.). The dimer formation rate $J_{dimer}$ versus SA monomer concentration is shown in Fig. 4. The dimer formation rates agree well with a simple analytical expression for the maximum possible NPF rate of SA particles (13) (solid line in Fig. 4). In comparison, the NPF rates at 1.7 nm are lower by 2–4 orders of magnitude (15). The difference between the NPF rates at these two sizes ($J_{dimer}$ vs. $J_{1.7nm}$) is due to losses during growth to 1.7 nm. Both the chamber walls and the clusters/particles act as condensation sinks for the growing clusters. The losses are largest at slow growth rates, corresponding to low SA concentration (37). In the atmosphere the condensation of vapors and clusters onto preexisting particles is similar to condensation on the walls of chamber experiments. In consequence the magnitude and slopes of experimental measurements of $J$ vs. $[H_2SO_4]$ at a chosen threshold size do not provide direct information on cluster evaporation or on the number of molecules in the critical cluster (37, 38). This is dramatically demonstrated in Fig. 4 because the SA dimer with DMA is already a stable particle, above the size of the critical cluster. For other systems where NPF is not proceeding at the kinetic limit, different cluster sizes would need to be chosen to determine the nucleation rates. Nevertheless, similar losses and slope distortions would occur between the critical size and the chosen particle threshold size. Regarding atmospheric nucleation, it is important to note that growth rates; condensation sinks; particle sizes, at which the formation rates have been determined; and probably also the chemical systems differ between different atmospheric observations. Therefore, it cannot be concluded that atmospheric nucleation is generally a kinetic process, although there is some overlap between the formation rates for the SA–DMA system and the boundary layer particle formation rates.

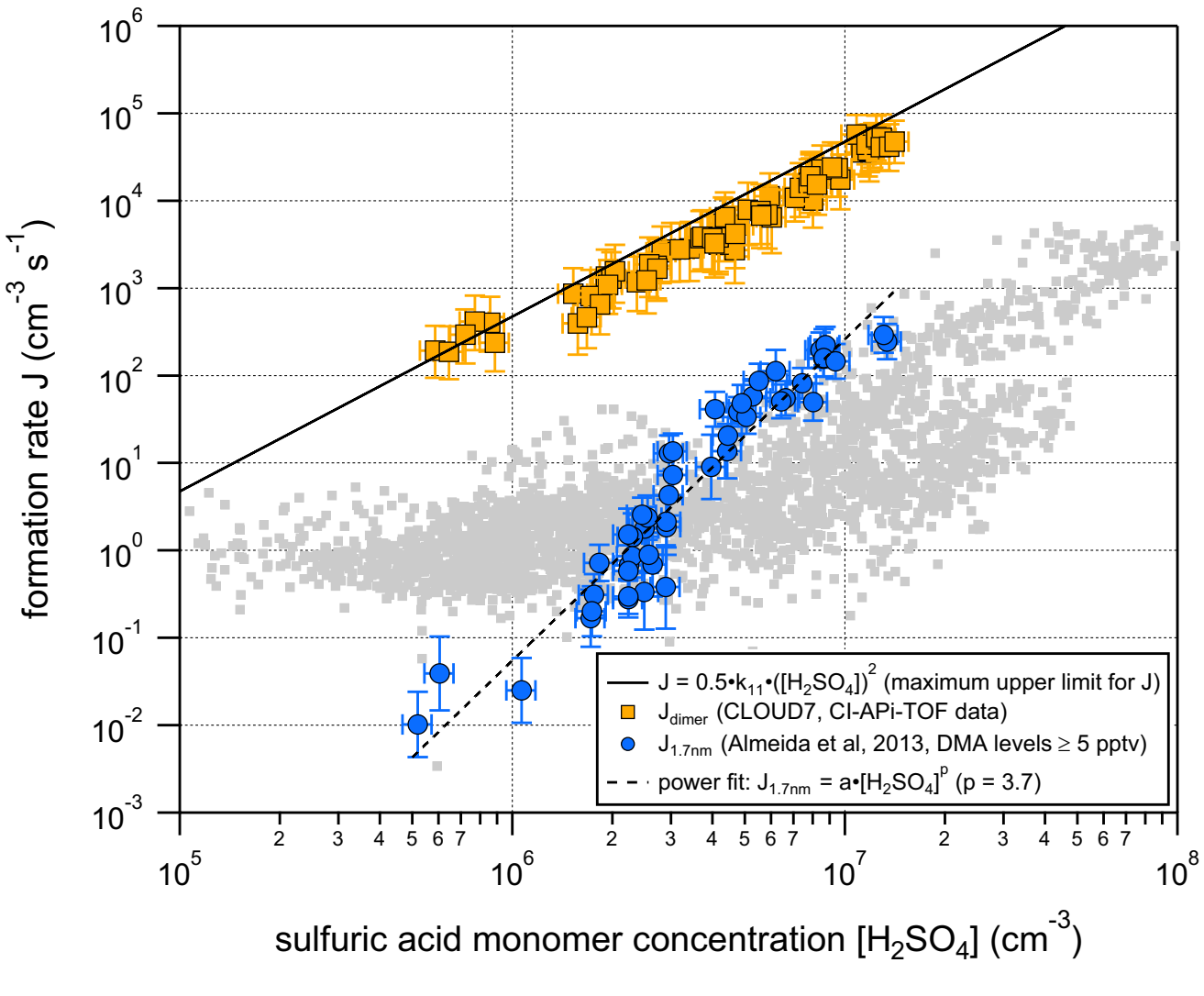

**Fig. 4.** Experimental, atmospheric, and theoretical particle formation rates against sulfuric acid monomer concentration (i.e., $[H_2SO_4]$ or $N_1$). Atmospheric boundary layer particle formation rates are shown by small gray squares (4, 6, 36). Formation rates for SA–DMA particles at a mobility diameter of 1.7 nm ($J_{1.7nm}$) previously measured by CLOUD at 5–140 pptv dimethylamine (DMA) are shown by the blue circles (15). These formation rates were derived using the standard method for calculating $J$ at a small size (39). A power law fit ($J = a \cdot [H_2SO_4]^p$) yields a slope of $p = 3.7$ for $J_{1.7nm}$ (dashed line). Neutral dimer formation rates from the present work are indicated by the orange squares ($J_{dimer}$). The maximum possible particle formation rate from kinetic sulfuric acid new particle formation is indicated by the solid curve ($J = 0.5 \cdot k_{11} \cdot [H_2SO_4]^2$), where $k_{11}$ is the collision rate between two sulfuric acid molecules (13). The difference between the kinetic dimer formation rate and the observed $J_{1.7nm}$ is almost entirely due to loss of the small clusters.

## Conclusions

Because atmospheric boundary layer nucleation is generally dominated by the neutral nucleation pathway, it is of utmost importance to study the formation of neutral clusters (6, 26, 27). Additionally, recent CLOUD chamber studies demonstrated that ion-induced nucleation is not substantial when the particle formation rates exceed ~1 $cm^{-3} \cdot s^{-1}$ in nucleating systems involving SA as well as DMA or oxidized organics (15, 25). In this study, the formation of neutral clusters containing up to 14 SA and 16 DMA molecules was observed for the first time to our knowledge during NPF in the CLOUD chamber, including their temporal evolution.

The formation of the neutral SA–DMA clusters follows a stoichiometry of very close to 1:1 largely independent of the investigated SA and DMA concentrations. This reveals that full neutralization of sulfuric acid with respect to its acidity does not occur at the observed cluster sizes because this would require a stoichiometry of 1:2 between acid and base.

We have shown that NPF of neutral SA–DMA clusters under atmospheric conditions proceeds at or near the kinetic limit, implying negligible evaporation, which is equivalent to the notion that the critical cluster size is smaller than the dimer. We find that the NPF rate of neutral SA dimers versus $[H_2SO_4]$ in the presence of DMA proceeds at the maximum rate expected for kinetically limited NPF, with a power dependence on $[H_2SO_4]$ of 2. However, due to particle losses, the formation rate at 1.7 nm is up to 4 orders of magnitude lower than the dimer formation rate and has a power dependence near 3.7. The implication is that the translation of experimental results into mechanisms appropriate for the atmosphere will require an understanding of the kinetics of NPF, growth, and loss from the first molecular collisions (39). Most importantly, cluster loss via collisions with larger particles, and not evaporation, can dominate for even the smallest clusters (40). We have developed a detailed understanding for the neutral SA–DMA system here, and similar progress now seems achievable for various other atmospherically relevant systems.

## Methods

The CLOUD chamber at CERN allows nucleation experiments to be conducted under exceptionally clean and well-defined conditions (5, 15). The 26.1-$m^3$ electropolished stainless steel chamber is filled with artificial air by mixing nitrogen and oxygen from cryogenic liquids at a ratio of 79:21. Additionally, $H_2O$, $O_3$, and $SO_2$ can be added; together with UV light, which is fed into the chamber by means of a sophisticated fiber-optic system, this allows the photolytic generation of $H_2SO_4$. Dimethylamine from a gas bottle diluted with nitrogen was fed into the chamber during the experiments studying the NPF of SA and DMA. Two magnetically driven mixing fans ensured the rapid distribution of DMA and the other trace gases. Contact between plastic materials and the gases flowing into the chamber is avoided to minimize the abundance of spurious compounds. The chamber temperature is precisely controlled (±0.01 K) and can be adjusted between 208 and 373 K. During all experiments discussed here, the temperature was 278 K, and the relative humidity was 38%. One of the main purposes of the CLOUD facility is to study

ENVIRONMENTAL SCIENCES

the influence of ions on nucleation and particle growth. Ions can be created by galactic cosmic rays and a pion beam from the CERN proton synchrotron. However, for the experiments shown here, an ion-free environment was required. Therefore, ±30 kV were applied to two opposing high-voltage field cage electrodes installed inside the CLOUD chamber. The strong electric field eliminates all ions within about 1 s and allows the neutral nucleation pathway to be studied. The data shown in this study were obtained by using the CI-APi-TOF (Tofwerk AG and Aerodyne Research) technique (30) (*SI Text*). The CI-APi-TOFs include a CI charging unit where nitrate ions $[NO_3^-(HNO_3)_{0-2}]$ are mixed with the gas sampled from the CLOUD chamber. The primary ions rapidly react with free $H_2SO_4$ molecules and SA–DMA clusters at ambient pressure and produce distinct product ions. Using the APi, the ions can be transferred from the reaction zone into a TOF mass spectrometer. Owing to the high mass resolving power and mass accuracy the elemental composition of the measured ions can be determined. The data were analyzed with the MATLAB-based software package tofTools (29). Dimethylamine mixing ratios were determined from measurements with an ion chromatograph (15, 41).

**ACKNOWLEDGMENTS.** We thank the European Center for Nuclear Research (CERN) for supporting Cosmics Leaving Outdoor Droplets (CLOUD) with important technical and financial resources and for providing a particle beam from the CERN Proton Synchrotron. We also thank P. Carrie, L.-P. De Menezes, J. Dumollard, K. Ivanova, F. Josa, I. Krasin, R. Kristic, A. Laassiri, O. S. Maksumov, B. Marichy, H. Martinati, S. V. Mizin, R. Sitals, A. Wasem, and M. Wilhelmsson for their important contributions to the experiment and P. H. McMurry for helpful discussion. We thank the tofTools team for providing tools for mass spectrometry analysis. This research has received funding from the European Commission Seventh Framework Programme (Marie Curie Initial Training Network "CLOUD-ITN" 215072, MC-ITN "CLOUD-TRAIN" 316662, European Research Council (ERC)-Starting "MOCAPAF" Grant 57360, and ERC-Advanced "ATMNUCLE" Grant 227463). This research has also received funding from the German Federal Ministry of Education and Research (Projects 01LK0902A and 01LK1222A), the Swiss National Science Foundation (Projects 200020 135307 and 206620 141278), the Academy of Finland (135054, 133872, 251427, 139656, 139995, 137749, 141217, 141451 and Center of Excellence Project 1118615), the Finnish Funding Agency for Technology and Innovation, the Väisälä Foundation, the Nessling Foundation, the Austrian Science Fund (Project J3198-N21), the Portuguese Foundation for Science and Technology (Project CERN/FP/116387/2010), the Swedish Research Council, Vetenskapsrådet (Grant 2011-5120), the Presidium of the Russian Academy of Sciences and Russian Foundation for Basic Research (Grants 08-02-91006-CERN and 12-02-91522-CERN), the US National Science Foundation (Grants AGS1136479 and CHE1012293), Pan-European Gas-Aerosols-Climate Interaction Study project [funded by the European Commission under the Framework Program 7 (FP7-ENV-2010-265148)], and the Davidow Foundation.

# Supporting Information

### SI Text

**CLOUD Chamber and Instruments.** The Cosmics Leaving Outdoor Droplets (CLOUD) experiment is designed to study the formation and growth of aerosol particles and the role of ions in these processes. Experiments were conducted at the CLOUD chamber in October 2012 for the ternary system involving sulfuric acid (SA), dimethylamine (DMA), and water vapor. The CLOUD experiment and the chamber have been described in previous publications (1, 2). A summary is given here that focuses on the aspects that are relevant for this study. The chamber consists of an electropolished stainless steel cylinder with a volume of 26.1 $m^3$. The neutral nucleation pathway can be studied when a high-voltage clearing field is enabled. This is achieved by applying +30 kV to an upper and −30 kV to a lower transparent field cage electrode. Ceramic spacers insulate the chamber from the electrodes, and the strong electric field sweeps out all ions produced by natural galactic cosmic rays within about 1 s. Ion-induced nucleation (IIN) is studied when the electrodes are grounded. In this case, natural galactic cosmic rays generate ions, which have been shown to enhance the new particle formation (NPF) rates for the binary sulfuric acid–water ($H_2SO_4$–$H_2O$) system as well as for the ternary system involving ammonia ($NH_3$) (1). For the ternary system with dimethylamine [$(CH_3)_2NH$], the ion-induced contribution is only important when sulfuric acid concentrations are low (2). Higher ionization rates are achieved by illuminating the chamber with a defocused charged pion beam from the European Organization for Nuclear Research (CERN) Proton Synchrotron. Ionization rates up to about 75 ion pairs·$cm^{-3}$·$s^{-1}$ can be reached. From the perspective of the neutral cluster detection it is necessary to distinguish between the neutral and the IIN case. During IIN studies it has been observed that charged clusters from the chamber can contribute to the product ion signals originating from the neutral clusters measured with the Chemical Ionization–Atmospheric Pressure interface–Time Of Flight (CI-APi-TOF) mass spectrometers. Although the number concentration of the charged clusters is much lower than for the neutral clusters, the charged clusters can still contribute to the total ion signal because only a small fraction of the neutral species is ionized within the CI-APi-TOF drift tube. Therefore, the CI-APi-TOF-U-FRA (instrument from the University of Frankfurt) employs an ion precipitator integrated in its sampling line to remove the chamber ions. The CI-APi-TOF-U-HEL (instrument from the University of Helsinki) was not equipped with an ion precipitator; therefore, the data shown for this instrument are limited to the neutral runs.

The CLOUD chamber has been designed to achieve a very high level of cleanliness by avoiding contact between plastic materials and the gas inside the chamber. Nitrogen and oxygen from cryogenic liquids, which should be free of contaminants, are used. Minimizing contaminants to the extent possible is necessary when nucleation studies are performed at atmospherically relevant sulfuric acid concentrations, between about $1 \times 10^6$ and $1 \times 10^7$ $cm^{-3}$, because contamination with amines at similar levels can substantially enhance the new particle formation rate (2). Therefore, it is necessary to monitor the contents of the chamber for such species. An ion chromatograph is used to determine the mixing ratio of dimethylamine and ammonia (3). The Proton-Transfer Reaction Time of Flight Mass Spectrometer (4) primarily monitors the concentrations of organic compounds (5), but it can also be used for the measurement of ammonia (6) or dimethylamine. The APi-TOF mass spectrometer identifies the molecular compositions of ions and cluster ions (7). The mass spectrometer is identical to the one that is also used in the CI-APi-TOFs and is described in the next section. The APi-TOF does not include a charging unit; therefore, it provides meaningful data only during experimental runs when ions are present. The presence of ammonia or amines associated with sulfuric acid clusters during a NPF event points to contamination in case these substances were not added intentionally. This is a very direct way of identifying compounds that are nucleating. At a temperature of 278 K, contaminant ammonia is present and can be detected in the APi-TOF mass spectra (1), but dimethylamine has not been detected during CLOUD7 when it was not added to the chamber. When present at sufficiently high concentrations, DMA will rapidly displace ammonia in the clusters as has been shown in previous experimental and theoretical studies (8, 9). The mass defect plot (Fig. 1*A*) shows that only minor amounts of $NH_3$ are present in the large clusters (heptamer and larger). Therefore, the data shown here are valid for the ternary system including only sulfuric acid, water and dimethylamine.

In addition to the two CI-APi-TOFs that are used to measure the sulfuric acid monomer and cluster concentrations (see next section), the Chemical Ionization Mass Spectrometer (CIMS) is used to determine the sulfuric acid monomer concentration (10–12).

Sulfuric acid production is initiated by UV light that is brought into the chamber through a fiber-optic system (13). Photolysis of ozone and subsequent reactions with water vapor, sulfur dioxide, and oxygen will generate sulfuric acid. When the $H_2SO_4$ loss rate equals its production rate, the concentration reaches a steady state. Depending on the UV light intensity and the trace gas mixing ratios, the concentration can be controlled. The gas is homogenously mixed by two fans installed inside the chamber (14). Although the sulfuric acid is produced in situ, dimethylamine is taken from a gas bottle. Before it is introduced into the chamber, it is diluted with clean air to achieve the desired mixing ratios. The addition of the diluted dimethylamine is performed close to the lower mixing fan, which ensures its rapid distribution throughout the chamber. The reported dimethylamine mixing ratios are from the IC measurement (2, 3).

**CI-APi-TOF Instruments.** The CI-APi-TOF technique has recently been described by Jokinen et al. (15). A schematic drawing of the CI-APi-TOF-U-FRA is shown in Fig. S1. Within the ion source, a corona discharge is used to initiate the formation of $NO_3^-(HNO_3)_n$ (usually $n \leq 2$) primary ions from nitric acid that is added to the sheath gas. The ion source and ion drift tube are an exact copy of the ion source used in the CIMS and have been described in detail by Kürten et al. (12).

The sample flow rate into the instrument is defined by the difference between the flow rates that are taken from the ion source and the drift tube, i.e., the excess air and the flow that enters the mass spectrometer through a small pinhole, and the flows that are actively introduced, i.e., the sheath gas (clean gas + $HNO_3$) and a flow of dry nitrogen in front of the pinhole. The sample flow rate is ~8.5 standard liters per minute. It enters the ion drift tube where it is surrounded concentrically by the sheath gas. The primary ions are directed toward the center of the sample flow by means of an electrostatic field so that they can interact with sulfuric acid monomers and clusters. These compounds can be ionized through proton-transfer reaction. The reaction scheme for the sulfuric acid monomer is

$$H_2SO_4 + NO_3^-(HNO_3)_n \rightarrow HSO_4^-(HNO_3)_{n-m+1} + m \cdot (HNO_3). \quad \textbf{[R1]}$$

The trajectory of the ions is defined by the geometry, flow field, and applied electrostatic voltages. Because the trajectories of the primary ions and, therefore, the effective reaction time are not known, it is necessary to calibrate the instrument with a known amount of sulfuric acid in the sample flow (16). The quantification of the sulfuric acid monomer and clusters is described in the next section. Primary and product ions enter the vacuum chamber through a small pinhole (~350 μm in diameter).

The mass spectrometer (Tofwerk AG) includes the electronics for the data acquisition as well as the software for controlling the instrument and recording the mass spectra. The vacuum chamber is separated into four chambers that are differentially pumped. The pressure in the first stage is ~3 hPa, which is maintained by a scroll pump (TriScroll 600; Agilent Technologies). A quadrupole mass filter (Quad1) is used as ion guide and helps to transfer the ions to the next chamber. The second chamber contains another quadrupole ion guide (Quad2) and is connected to the first stage of a three-stage turbo pump. The third stage contains a lens stack that is used to focus the ions and to prepare them energetically before they reach the final stage consisting of the time-of-flight mass spectrometer. These last two stages are also connected to the three-stage turbo pump. The pressure in the time-of-flight chamber is $\sim 1 \times 10^{-6}$ hPa. A high-voltage pulse is used to deflect the ions and accelerate them toward a reflectron. The mass spectrometer can either be operated using just one reflectron (so-called V-mode due to the shape of the ion trajectories), or a second reflectron can be used to increase the mass resolving power through a longer flight path (so-called W-mode). Because the high mass resolving power can only be achieved at the expense of a reduced sensitivity, the V-mode was used throughout this study for both CI-APi-TOFs. Detection of the ions is achieved with a multichannel plate detector. The vacuum chamber part is also used in the APi-TOF mass spectrometer, which has been described in detail by Junninen et al. (7). Typically, preaveraged mass spectra are recorded with a time resolution of 5 s. For the evaluation of the time-of flight mass spectra the MATLAB-based Toftools software is used (7). The CI-APi-TOFs usually achieve a mass accuracy of better than 10 ppm and a mass resolving power up to 4,500 Th/Th.

The CI-APi-TOF-U-HEL and the CI-APi-TOF-U-FRA differ in certain aspects from each other. Although the originally developed CI-APi-TOF-U-HEL instrument used a radioactive $^{241}$Am ion source, this source could not be used during the CLOUD experiment due to CERN's strict safety regulations. Therefore, an alternative method was deployed which makes use of a soft X-ray source (soft X-ray tube, N7599; Hamamatsu Photonics K.K.). The X-ray source is located outside of the ion source flange; the radiation is transmitted into the annular gap (where the corona needle is located in the CI-APi-TOF-U-FRA instrument in Fig. S1) through a thin Teflon foil. The interaction of the soft X-rays with the $HNO_3$ containing sheath gas produces the nitrate primary ions. This method yields very clean spectra with a stable ion count rate. The two instruments also differ in their sample tube diameters, drift tube lengths, and inner diameters. The relatively large dimensions of the Helsinki instrument result in an effective reaction time of ~200 ms; that of the smaller Frankfurt instrument is on the order of 50 ms. The different reaction times are taken into account by calibrating each instrument individually.

The corona discharge was found to lead to a greater abundance of background peaks than did the X-ray source. Increasing the amount of $HNO_3$ added to the sheath gas reduced the intensities of these signals. For this reason the amount of $HNO_3$ used in the sheath gas of the CI-APi-TOF-U-FRA was higher than in the Helsinki instrument. Because the neutral nitric acid can also interact with the sample gas, the increased $HNO_3$ concentration produces more SA–DMA clusters associated with nitric acid. However, although the spectra of the two instruments differ somewhat in this aspect, they show qualitatively the same results. In addition, deriving the cluster concentrations by summing up all signals related to a certain number of sulfuric acid molecules contained in the clusters yields remarkably good agreement between the two instruments (Fig. 3). This indicates that clustering with nitrate does not significantly influence the cluster detection efficiency.

As mentioned above, the CI-APi-TOF-U-FRA uses an ion precipitator integrated in its sampling line. The ion precipitator consists of a small piece of 0.5-inch stainless steel tubing which has been cut into two halves in the direction of flow. Applying 2 kV on one side and ground potential on the other half effectively removes all ions from the sample flow during ion-induced nucleation experiments.

A flow of nitrogen is added in front of the pinhole of the Frankfurt instrument (Fig. S1). The same design is used in CIMS instruments for the measurement of the sulfuric acid concentration and has, therefore, been adopted also for the CI-APi-TOF-U-FRA. When the ions travel through the dry nitrogen, water molecules are effectively removed from the core ions (17). This simplifies the mass spectra by avoiding that signals corresponding to a cluster with a given amount of sulfuric acid molecules are distributed over many peaks due to different numbers of water molecules associated with the core ion. Moreover, the nitrogen counterflow prevents the entry of fine particles and nitric acid into the vacuum chamber. The CI-APi-TOF-U-HEL does not use the $N_2$ counterflow.

Fragmentation of clusters as they transit from ambient pressure into the ultrahigh vacuum of the mass spectrometer cannot be ruled out (2, 18). This most likely happens in the Quad1 region where ion acceleration leads to energetic collisions with neutrals at relatively high pressure (several hPa). The extent of fragmentation is not known and needs to be further investigated in future studies. For this study, it can be concluded, however, that any fragmentation should affect mainly the trimer and the larger clusters. The agreement between modeled and measured dimer concentration is quite good (Figs. 2 and 3); moreover, the binding energy of the dimer ion $[HSO_4^-(H_2SO_4)]$ is very high, which should prevent its fragmentation when the CI-APi-TOFs are tuned to maximize the ratio between $HSO_4^-(HNO_3)$ and $HSO_4^-$ (19). If a large fraction of the sulfuric acid monomer is detected as $HSO_4^-(HNO_3)$, the dimer should not fragment substantially because the binding energy of $HSO_4^-(HNO_3)$ (27.4 kcal·mole$^{-1}$) is considerably lower than for $HSO_4^-(H_2SO_4)$ (41.8 kcal·mole$^{-1}$) (20, 21). Fragmentation of clusters larger than the dimer could be occurring to some extent. This does, however, not change the interpretation from Fig. 3 that NPF is very likely proceeding at the kinetic limit. Fragmentation should reduce all measured cluster concentrations at a certain size by a constant factor. Because the slope of $N_{cluster}$ vs. $N_1$ agrees best with the model calculations assuming zero evaporation, it can be concluded that the reduction in the cluster concentrations for $N_3$ and the larger clusters compared with the modeled concentrations are not due to evaporation of the neutral clusters.

**Cluster Quantification.** The sulfuric acid monomer concentration is estimated to be

$$[H_2SO_4] = N_1 = \frac{C_1}{T_1} \cdot \frac{S_{97} + S_{160}}{S_{62} + S_{125} + S_{188}}; \qquad \textbf{[S1]}$$

that is, it is proportional to the sum of the product ion signals $S_{97}$ ($m/z$ 97, $HSO_4^-$) and $S_{160}$ [$m/z$ 160, $HSO_4^-(HNO_3)$] divided by the sum of the primary ion signals $S_{62}$ ($m/z$ 62, $NO_3^-$), $S_{125}$ [$m/z$ 125, $NO_3^-(HNO_3)$], and $S_{188}$ [$m/z$ 188, $NO_3^-(HNO_3)_2$]. The constant $C_1$ is derived from calibration of the CI-APi-TOFs, during which a known concentration of $H_2SO_4$ is generated and from the measured signals the calibration constant is derived by the method described in ref. 16. The transmission efficiency of the sulfuric acid monomer through the sampling line from the CLOUD chamber to the ion drift tube of the CI-APi-TOFs is taken into account by the factor $T_1$. For straight circular tubes and laminar flow the transmission can be calculated from empirical equations (22). However, the two CI-APi-TOFs were connected to the CLOUD chamber with one common sampling line which was split after a certain distance connecting each instrument to one arm of the y-splitter. For this geometry the transmission cannot be calculated with empirical equations. Therefore, the transmission efficiency was derived from comparison of the measured sulfuric acid monomer concentration with the CIMS (see above). This instrument has its own sampling line which consists of a straight tube and the CIMS was calibrated individually with the same calibration system as the CI-APi-TOFs. The transmission efficiency $T_1$ that has been derived from this method has a value of 0.32.

The evaluation of the cluster concentrations is more difficult because it is not yet possible to calibrate for these species. Generally, their concentrations can be derived from the following formula:

$$N_i = \frac{C_1}{T_1} \cdot \frac{k_1}{k_i} \cdot \frac{T_1}{T_i} \cdot \frac{e_1}{e_i} \cdot \frac{\sum \text{product ion signals}}{S_{62} + S_{125} + S_{188}}. \qquad \textbf{[S2]}$$

This equation takes into account three effects which lead to differences from the monomer. The first effect (term $k_1/k_i$) is the different reaction rate between the cluster and the primary ions compared with the monomer. Therefore, the equation needs to be scaled with the monomer reaction rate divided by the cluster reaction rate. The values for the monomer and the clusters are $1.9 \times 10^{-9}$ $cm^3 \cdot s^{-1}$ ($k_1$ and $k_2$) and $2.2 \times 10^{-9}$ $cm^3 \cdot s^{-1}$ ($k_3$ to $k_5$) [$k_1$ to $k_4$ from Chen et al. (19); $k_5$ tentatively set to the same value as $k_4$]. It should be noted that these rate constants are derived for pure sulfuric acid clusters. The presence of DMA in the clusters will likely change the ionization efficiencies depending on the amount of DMA. Currently, it is, however, not possible to calibrate for this effect. Therefore, the above values are being used.

The second correction term $T_1/T_i$ accounts for the increase in the transmission efficiency through the sampling line with increasing size of the molecule or cluster due to its smaller diffusivity. This effect has been quantified by deriving an effective length for the known monomer transmission efficiency (23). Because this efficiency is known as a function of the flow rate and the diffusivity of the monomer, the length for which the experimentally determined monomer transmission efficiency would result can be calculated. With this information the transmission $T_i$ for the clusters can be calculated with the equations provided in ref. 22.

The third term, $e_1/e_i$, considers mass discrimination effects from the acceptance of the pinhole, the quadrupole ion guides (Quad1 and Quad2), the time-of-flight mass spectrometer, and the multichannel plate detector. The relevant mass range of the product ions spans 97 Th ($HSO_4^-$ ion) to 776 Th [$HSO_4^-(H_2SO_4)_4((CH_3)_2NH)_5(HNO_3)$ ion], and therefore, differences in the detection efficiency between light and heavy ions can be expected. However, it is not trivial to quantify this effect. This was attempted by calibration experiments where ions were generated by electrospray. Subsequently, ions within a narrow mobility range were selected with a high-resolution differential mobility analyzer (24). The flow from the high-resolution differential mobility analyzer was split, and one part was fed into an electrometer, whereas the other one was used as the sample flow for the CI-APi-TOF. The ratio of the signals obtained from the CI-APi-TOF and the electrometer for different ionic species covering a wide range of $m/z$ values allows us to derive a relative transmission efficiency curve for the CI-APi-TOF. This is possible because the electrometer has a detection efficiency that is effectively independent of the ion mass if multiple charges can be ruled out. However, the resulting transmission curve cannot be directly applied to neutral clusters, because a fraction of the ions is precipitated in the sampling line before they can enter the drift tube during this calibration procedure. The negative ions experience a repulsing electric field just before they are transferred from the sampling line into the drift tube and are, therefore, accelerated toward the walls of the sampling line. This effect depends strongly on the ion mobility and, therefore, affects the small ions to a larger extent than the heavier ions.

For this reason, no corrections according to the obtained transmission curves were applied. Instead, a different method was used to verify that the monomer and dimer concentrations show similar transmission efficiencies. During the CI-APi-TOF calibration, high sulfuric acid monomer concentrations were generated. Under the clean conditions during a calibration, the neutral dimer concentration is negligible. Therefore, if the signal at $m/z$ 195 [$HSO_4^-(H_2SO_4)$] is elevated, it is due to ion clustering between $HSO_4^-$ product ions and $H_2SO_4$ within the CI-APi-TOF drift tube (25). The expected $m/z$ 195 signal due to this process is (26)

$$S_{195} = \frac{1}{2 \cdot C_1^2} \cdot [H_2SO_4]^2 \cdot (S_{62} + S_{125} + S_{188}). \qquad \textbf{[S3]}$$

Here [$H_2SO_4$] is the applied sulfuric acid monomer concentration and $C_1$ is the calibration constant for the monomers. Good agreement between the expected and the measured $S_{195}$ indicates that the detection efficiency for the monomer and the dimer is very similar. Therefore, although the exact quantification of the trimer and larger clusters is not possible at the moment, the dimer concentration can be reported with a higher confidence. It should be noted that the above discussion (and also Eq. **S2**) leaves out the effect of potential cluster fragmentation.

Note that little ion clustering occurs in the CI drift tube during the NPF experiments because the sulfuric acid concentration is low enough (and the reaction/residence time is short enough) to prevent this effect.

**Kinetic Model.** The kinetic model that is used to calculate the cluster distributions is based on ref. 27. The time-dependent balance equation for the monomer concentration $N_1$ is

$$\frac{dN_1}{dt} = P_1 - \left(k_{1,w} + k_{dil} + \sum_{j=1}^{N} G_{1,j} \cdot \beta_{1,j} \cdot N_j\right) \cdot N_1 + 2 \cdot k_{2,evap} \cdot N_2. \qquad \textbf{[S4]}$$

For the dimer the time-dependent concentrations can be calculated by

$$\frac{dN_2}{dt} = \frac{1}{2} \cdot G_{1,1} \cdot \beta_{1,1} \cdot N_1 \cdot N_1 - \left(k_{2,w} + k_{dil} + \sum_{j=1}^{N} G_{2,j} \cdot \beta_{2,j} \cdot N_j\right) \cdot N_2 - k_{2,evap} \cdot N_2, \qquad \textbf{[S5]}$$

whereas for all larger clusters ($k > 2$),

$$\frac{dN_k}{dt} = \frac{1}{2} \cdot \sum_{i+j=k} G_{i,j} \cdot \beta_{i,j} \cdot N_i \cdot N_j - \left(k_{k,w} + k_{dil} + \sum_{j=1}^{N} G_{k,j} \cdot \beta_{k,j} \cdot N_j\right) \cdot N_k. \qquad \textbf{[S6]}$$

Here $P_1$ is the production rate of the monomers due to the generation of OH after the photolysis of ozone and subsequent reactions with water vapor, sulfur dioxide, and oxygen. The model of McMurry (27) has been extended to include the dimer evaporation rate ($k_{2,evap}$). All larger clusters are assumed to be stable. The loss terms in Eqs. **S4**–**S6** include the wall loss rate $k_{k,w}$ and the dilution rate $k_{dil}$ that results from replenishment of the gas sampled by the instruments with clean gas. The dilution rate $k_{dil}$ equals $9.6 \times 10^{-5}$ $s^{-1}$ and is determined by the ratio of the clean gas flow rate into the chamber (150 standard liters per minute) and the chamber volume (26.1 $m^3$). This factor is independent of the cluster size, whereas the wall loss rate depends on the diffusivity of the molecule or cluster (28, 29):

$$k_{k,w} = C_w \cdot \sqrt{D_k}. \quad \textbf{[S7]}$$

The prefactor $C_w$ has been estimated from experiments in which the decrease in the sulfuric acid or particle concentration has been observed as a function of time, $C_w = 0.0077\ \mathrm{cm}^{-1}\cdot\mathrm{s}^{-0.5}$. The diffusivity is calculated as function of the molecular weight of the cluster, temperature, and pressure. The third loss term describes the depletion of monomers due to self-coagulation and coagulation with larger clusters. The coagulation coefficient β is derived from kinetic theory (30),

$$\beta_{k,j} = \left(\frac{3}{4\pi}\right)^{1/6} \cdot \sqrt{\left(\frac{6k_bT}{m_k} + \frac{6k_bT}{m_j}\right)} \cdot \left(V_k^{1/3} + V_j^{1/3}\right)^2. \quad \textbf{[S8]}$$

It depends on the temperature $T$, the masses $m_k$ and $m_j$ of the clusters $k$ and $j$, and their respective volumes $V_k$ and $V_j$; $k_b$ is the Boltzmann constant. The factor $G_{k,j}$ expresses the enhancement in the collision rates due to London–van der Waals forces and can be calculated from the formulas and the Hamaker constant given in ref. 31. The evaluated factors $G_{i,j}$ are around 2.3 for the free molecule regime, which is close to the value reported for nanometer-sized ammonium sulfate particles (32).

For simplicity it has been assumed that the clusters $k$ are of the form $(H_2SO_4)_k((CH_3)_2NH)_k$. This means that when the concentration of dimethylamine in the presence of sulfuric acid is sufficiently high, all sulfuric acid is associated with DMA. This assumption is in accordance with quantum chemical calculations which suggest that clusters containing equal amounts of SA and DMA have very low evaporation rates (2, 30, 33, 34). Nevertheless, these calculations show that the evaporation rate of the smallest SA•DMA clusters is still nonnegligible. However, if the DMA concentration is large enough, the SA•DMA clusters form rapidly, and their fraction is large compared with the overall sulfuric acid concentration (sum of the free SA molecules and the SA•DMA clusters) (2). In this respect, large enough means that the arrival rate of a DMA molecule on a sulfuric acid molecule is at least as fast as the evaporation rate of a SA•DMA cluster. Ortega et al. (30) report an evaporation rate of $5.9 \times 10^{-2}\ \mathrm{s}^{-1}$. Using the DMA mixing ratios during the experiments (between 5 and 32 pptv, i.e., concentrations between $1.3 \times 10^8$ and $8.3 \times 10^8\ \mathrm{cm}^{-3}$) and a collision rate $k_{SA,DMA}$ of $5 \times 10^{-10}\ \mathrm{cm}^3\cdot\mathrm{s}^{-1}$, the arrival rate of a DMA molecule on a SA molecule can be calculated to be between $6.5 \times 10^{-2}$ and $0.42\ \mathrm{s}^{-1}$. Because these values are larger than the evaporation rate, it is justified to treat SA•DMA clusters as a single molecule in the kinetic model. Because the evaporation rates were reported for a temperature of 298 K (30) and the experiments were conducted at 278 K in this study, the stated evaporation rate is an upper limit.

The volumes in Eq. **S8** require the knowledge of the cluster densities. The density of the clusters is determined as the weighted average of the liquid bulk densities of sulfuric acid ($1.84\ \mathrm{g\cdot cm^{-3}}$) and dimethylamine ($0.67\ \mathrm{g\cdot cm^{-3}}$). This yields a density of $1.47\ \mathrm{g\cdot cm^{-3}}$. Fission, i.e., nonmonomer evaporation from neutral clusters, was predicted based on quantum chemical calculations, e.g., for the cluster containing four SA and four DMA molecules (30, 34). However, the tetramer fission rate of $5 \times 10^{-2}\ \mathrm{s}^{-1}$ is rather low (34). From our experimental results it cannot be concluded whether the fission of neutral tetramers is indeed occurring. If this would be the case the fission rate would probably be even lower than reported; otherwise, the slope of $N_4$ vs. $N_1$ in Fig. 3*C* should be steeper.

For the time-dependent cluster concentration modeling (Fig. 2) the production term $P_1$ in Eq, **S4** is adjusted until the modeled steady-state monomer concentration $N_1$ matches the measured concentration. The model results shown in Fig. 3 are obtained by varying the monomer concentration over the range from $1 \times 10^6$ to $2 \times 10^7\ \mathrm{cm}^{-3}$. For each model run the cluster concentrations are calculated until a steady-state is reached. For the results shown, clusters up to $k$ = 2,000 ($d_{mob}$ ~9 nm) are included. For the accuracy of the model results it is actually not necessary to include that many clusters because the loss rate due to coagulation with the very large clusters is generally negligible in comparison with wall loss and loss to the smallest clusters.

**Calculation of Dimer Formation Rates.** In steady-state the production rate ($P_2$) or formation rate of the dimers ($J_{dimer}$) equals their loss rate ($L_2$):

$$\frac{dN_2}{dt} = P_2 - L_2 = J_{dimer} - L_2 = 0.$$

Dimer formation rates from the CI-APi-TOFs in Fig. 4 were therefore determined from the overall loss rate ($L_2$) and the steady-state dimer concentration ($N_2$),

$$J_{dimer}(t) = N_2 \cdot \left(\overline{CS}_2 + k_{2,w} + k_{dil}\right). \quad \textbf{[S9]}$$

Here it is taken into account that the dimers are lost due to coagulation, wall loss, and dilution of the chamber gas. The coagulation sink of the dimer can be determined from

$$\overline{CS}_2 = \overline{CS}_{2,\text{CI-APi-TOF}} + \overline{CS}_{2,\text{PSM}} + \overline{CS}_{2,\text{SMPS}}. \quad \textbf{[S10]}$$

The contributions of the coagulation sink for different cluster/particle size ranges were calculated based on the measurements from three different instruments. The first term takes into account the coagulation of dimers due to self-coagulation and coagulation with clusters up to the pentamer,

$$\overline{CS}_{2,\text{CI-APi-TOF}} = \sum_{k=1}^{5}\left(G_{2,k} \cdot \beta_{2,k} \cdot N_k\right). \quad \textbf{[S11]}$$

The second term considers the loss of dimers on small particles. The number density of particles in the size range between 1.3 and 3 nm was measured by the particle size magnifier (PSM) (35), operating in scanning mode (36):

$$\overline{CS}_{2,\text{PSM}} = \sum_{k=d_{p,1}}^{d_{p,N}}\left(G_{2,k} \cdot \beta_{2,k} \cdot N_k\right). \quad \textbf{[S12]}$$

Loss on larger particles was taken into account by using the size distributions obtained with a scanning mobility particle sizer (SMPS) starting at diameters around 4 nm:

$$\overline{CS}_{2,\text{SMPS}} = \sum_{k=d_{p,1}}^{d_{p,N}}\left(G_{2,k} \cdot \beta_{2,k} \cdot N_k\right). \quad \textbf{[S13]}$$

The average of a time period of a nucleation run where the dimer formation rates reach a steady-state determines the reported $J_{dimer}$ in Fig. 4.

**Comparison Between Measured Data and Model Results.** To find out which model curve from Fig. 3 best describes the measured data, the ratio of the measured and the modeled concentrations was calculated for all data according to

$$r_i\left(k_{2,evap}\right) = \frac{N_{i,measured}}{N_{i,modeled}\left(k_{2,evap}\right)}, \quad \textbf{[S14]}$$

where $i = 2, 3, 4$, and 5 and $k_{2,evap}$ is either 0, $10^{-2}$, or $10^{-1}$ $s^{-1}$. The results for zero dimer evaporation are shown in Fig. S2. This figure shows the factor, which best describes the discrepancy between modeled and measured data. There is a clear trend that the larger clusters are detected with a lower efficiency. The main cause for this effect needs to be investigated in the future. It is, however, suspected that mass discrimination in the mass spectrometer and charging efficiency play the most important role.

The results taking into account different evaporation rates in the model and the CI-APi-TOF-U-FRA data are shown in Fig. S3. From this figure it is evident that the ratios calculated for $k_{2,evap} = 0$ $s^{-1}$ yield the most consistent values with the smallest scatter (SD). This supports the assumption that the deviation between measured and modeled data can be explained by a constant scaling factor, which arises from the uncertainties in the charging and the detection efficiencies of the clusters.

To test whether random variation can be responsible for the deviation from a constant ratio, statistical tests ($f$ test) have been performed, which test the validity of the following zero hypotheses:

$$\text{a)}\quad \mathrm{Var}\left[r_i\left(k_{2,\mathrm{evap}}=0\right)\right]=\mathrm{Var}\left[r_i\left(k_{2,\mathrm{evap}}=10^{-2}\mathrm{s}^{-1}\right)\right] \qquad \textbf{[S15]}$$

and

$$\text{b)}\quad \mathrm{Var}\left[r_i\left(k_{2,\mathrm{evap}}=0\right)\right]=\mathrm{Var}\left[r_i\left(k_{2,\mathrm{evap}}=10^{-1}\mathrm{s}^{-1}\right)\right] \qquad \textbf{[S16]}$$

for each cluster $i = 2, 3, 4$, and 5. The $f$ test yields a $P$ value describing the probability to obtain the given samples if the zero hypothesis were correct. Therefore, low values indicate that it is quite improbable that the zero hypothesis is correct. The test results ($P$ values) are given in the annotations of Fig. S3. For example, a value of $1.33 \times 10^{-15}$ (Fig. S3*A*) indicates that the hypothesis "the variations in $r_2(k_{2,evap} = 0\ s^{-1})$ are identical to the variations in $r_2(k_{2,evap} = 10^{-2}\ s^{-1})$" is correct only with an extremely low probability of $1.33 \times 10^{-15}$. Testing for similarity between the SDs of $r_2(k_{2,evap} = 0\ s^{-1})$ and $r_2(k_{2,evap} = 10^{-1}\ s^{-1})$ yields a probability of zero. From this perspective it is very likely that the dimer evaporation rates are smaller than $10^{-2}$ $s^{-1}$. Performing the same analysis for the Helsinki data (CI-APi-TOF-U-HEL data from Fig. 3) yields the result that the dimer evaporation rates are smaller than 0.1 $s^{-1}$.

**Further Evidence for Clusters Forming at the Kinetic Limit.** Independent evidence that indicates absence of significant cluster evaporation is provided by the time development of the clusters at the start of the run. Fig. S4 shows cluster concentrations ($N_1$ to $N_5$) recorded during a nucleation experiment where the monomer concentration ($N_1$) reached a maximum value of $2.2 \times 10^6$ $cm^{-3}$ during steady-state. Similarly, the cluster concentrations ($N_2$ to $N_5$) reached a constant value. Normalizing all cluster concentrations by their respective steady-state values yields the experimental data shown in Fig. S4. The same normalization was performed for the calculated cluster concentrations from the kinetic model (solid lines in Fig. S4). This allows the time development of the modeled and measured cluster concentrations to be compared without making any assumptions on the detection efficiency of the CI-APi-TOF mass spectrometer. Assuming an evaporation rate of zero for the dimer (Fig. S4*A* with $k_{2,evap} = 0$ $s^{-1}$) yields good agreement between measured and modeled appearance times of the clusters. Introducing finite evaporation rates of 0.01 $s^{-1}$ or 0.1 $s^{-1}$ (Fig. S4 *B* and *C*, respectively) predicts slower appearance times of the clusters that are incompatible with experimental measurements. The comparison between measured and modeled normalized cluster concentrations also reveals that the theoretical collision rates adequately describe the cluster dynamics.

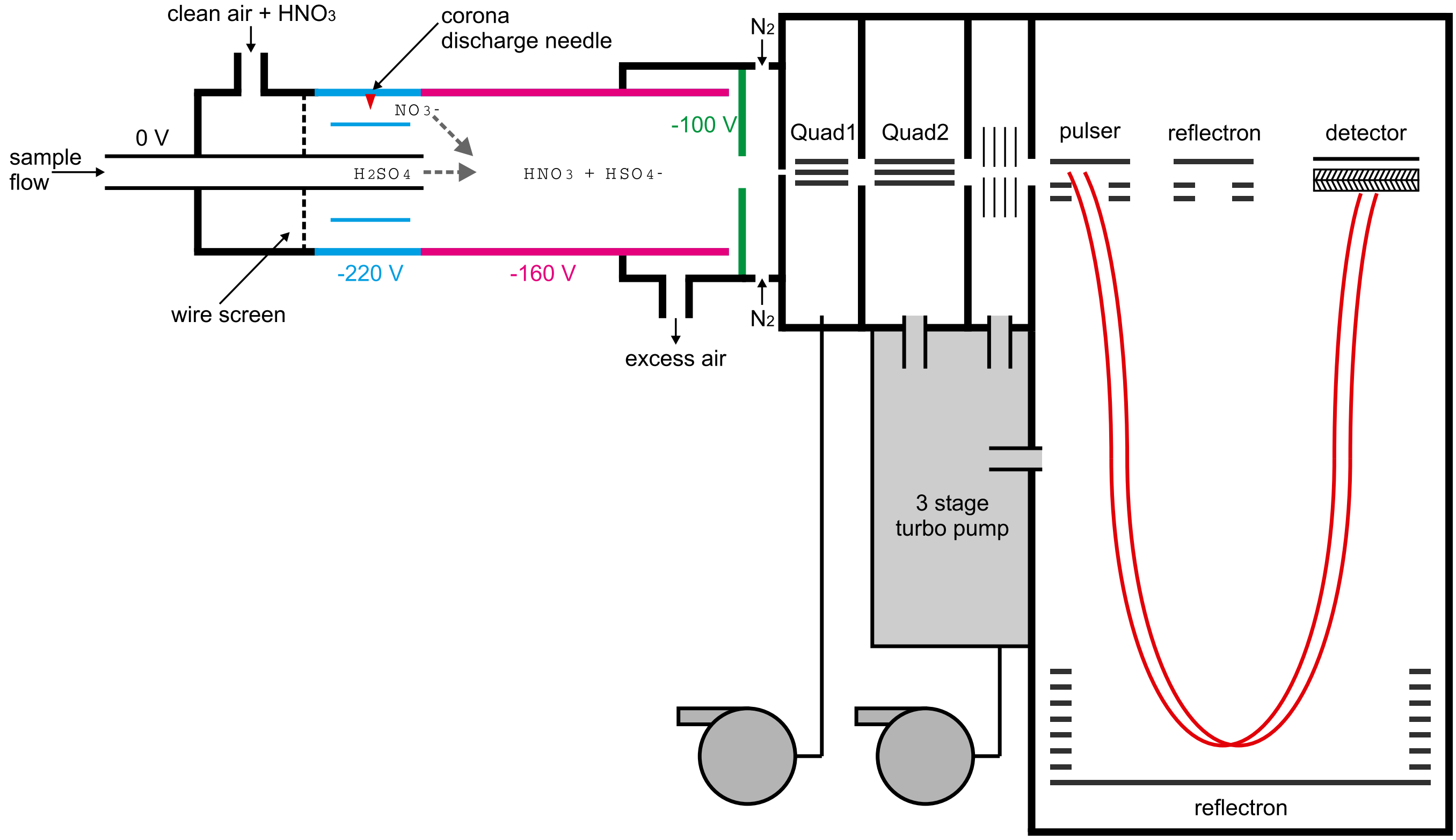

**Fig. S1.** Schematic drawing of the CI-APi-TOF mass spectrometer. Drawing is showing the instrument from the University of Frankfurt (CI-APi-TOF-U-FRA), which uses a corona discharge to generate the primary ions. The instrument from the University of Helsinki (CI-APi-TOF-U-HEL) uses a soft X-ray source for this task. The two instruments also differ in certain other details (*SI Text*). Drawing is not to scale.

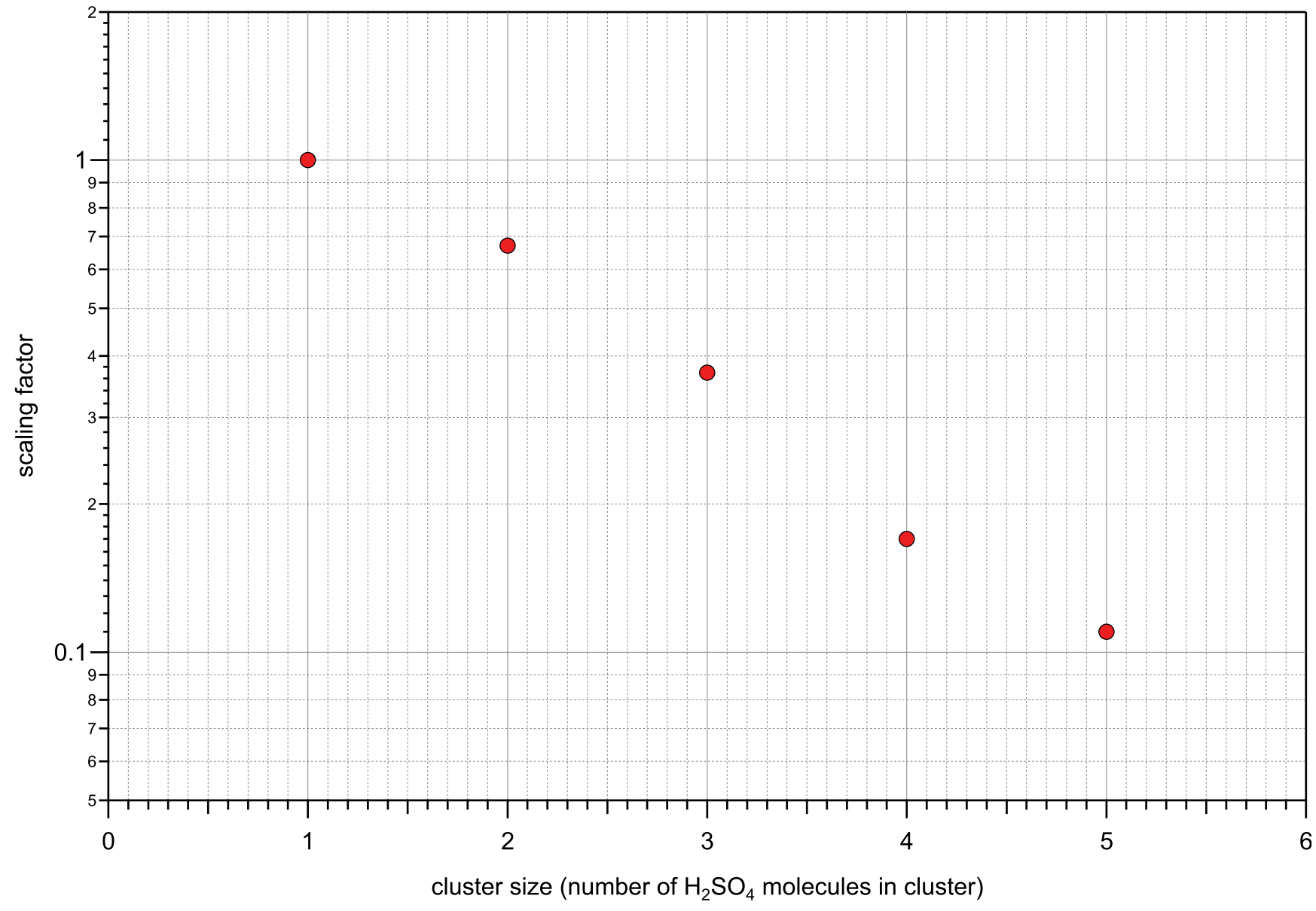

**Fig. S2.** Scaling factor for different cluster sizes. Scaling factor is derived by dividing the measured cluster concentrations by the modeled concentrations assuming zero dimer evaporation (Fig. 3). Qualitatively, the smooth decrease in the scaling factor is consistent with a decrease in the mass spectrometer sensitivity rather than with an increase in cluster evaporation rates.

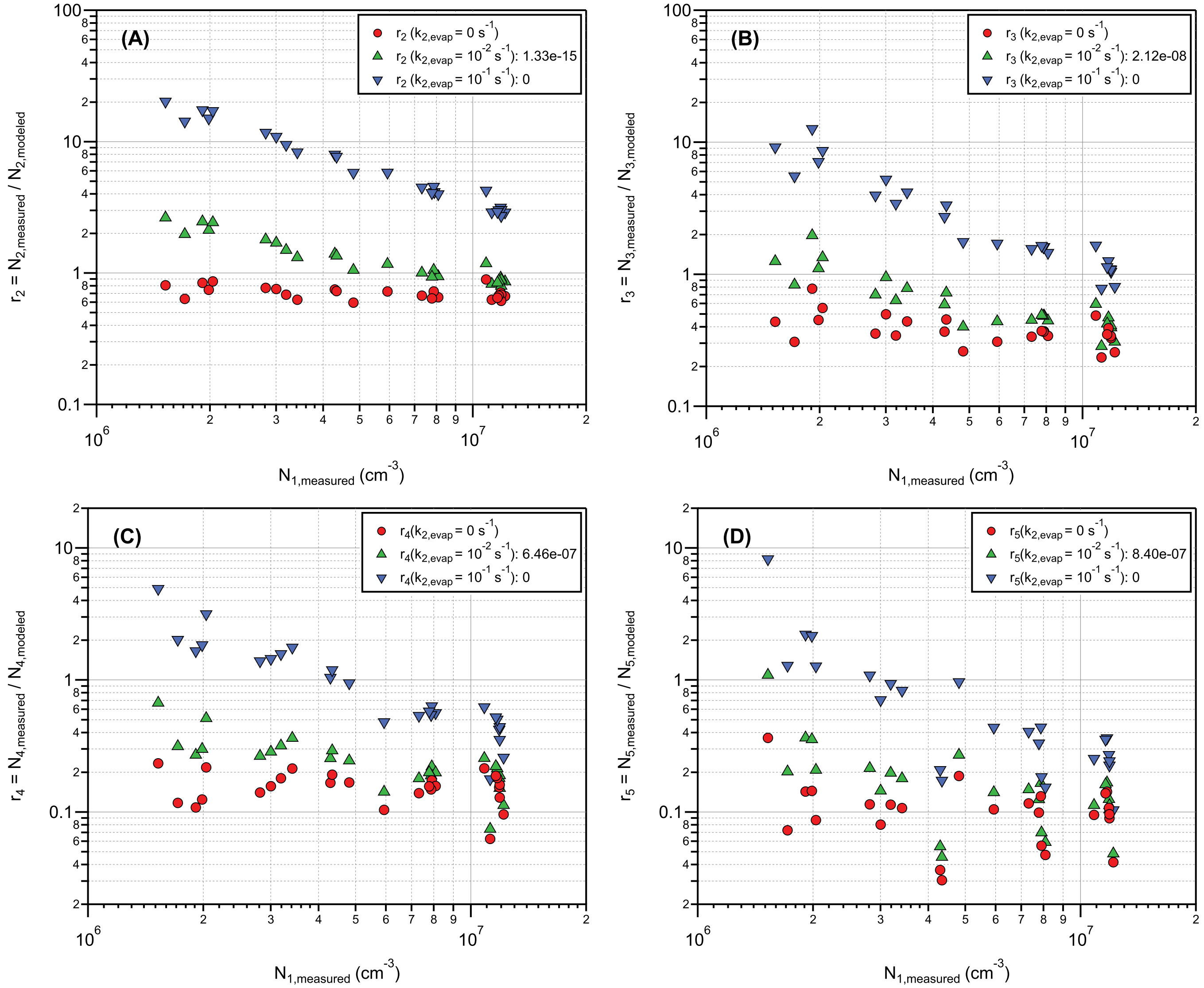

**Fig. S3.** Ratios between measured and calculated cluster concentrations as function of the sulfuric acid monomer concentration. Results are shown for (*A*) the dimer, (*B*) the trimer, (*C*) the tetramer, and (*D*) the pentamer for the measured data by the CI-APi-TOF mass spectrometer from the University of Frankfurt (CI-APi-TOF-U-FRA). Assuming different evaporation rates in the model, three different ratios were calculated for each cluster size. The numbers in the figure legend provide information about the similarity between the variance of the ratios assuming no dimer evaporation ($k_{2,evap} = 0\ s^{-1}$) and the variance of the ratios assuming nonzero dimer evaporation ($k_{2,evap} \geq 10^{-2}\ s^{-1}$). See *SI Text* for details.

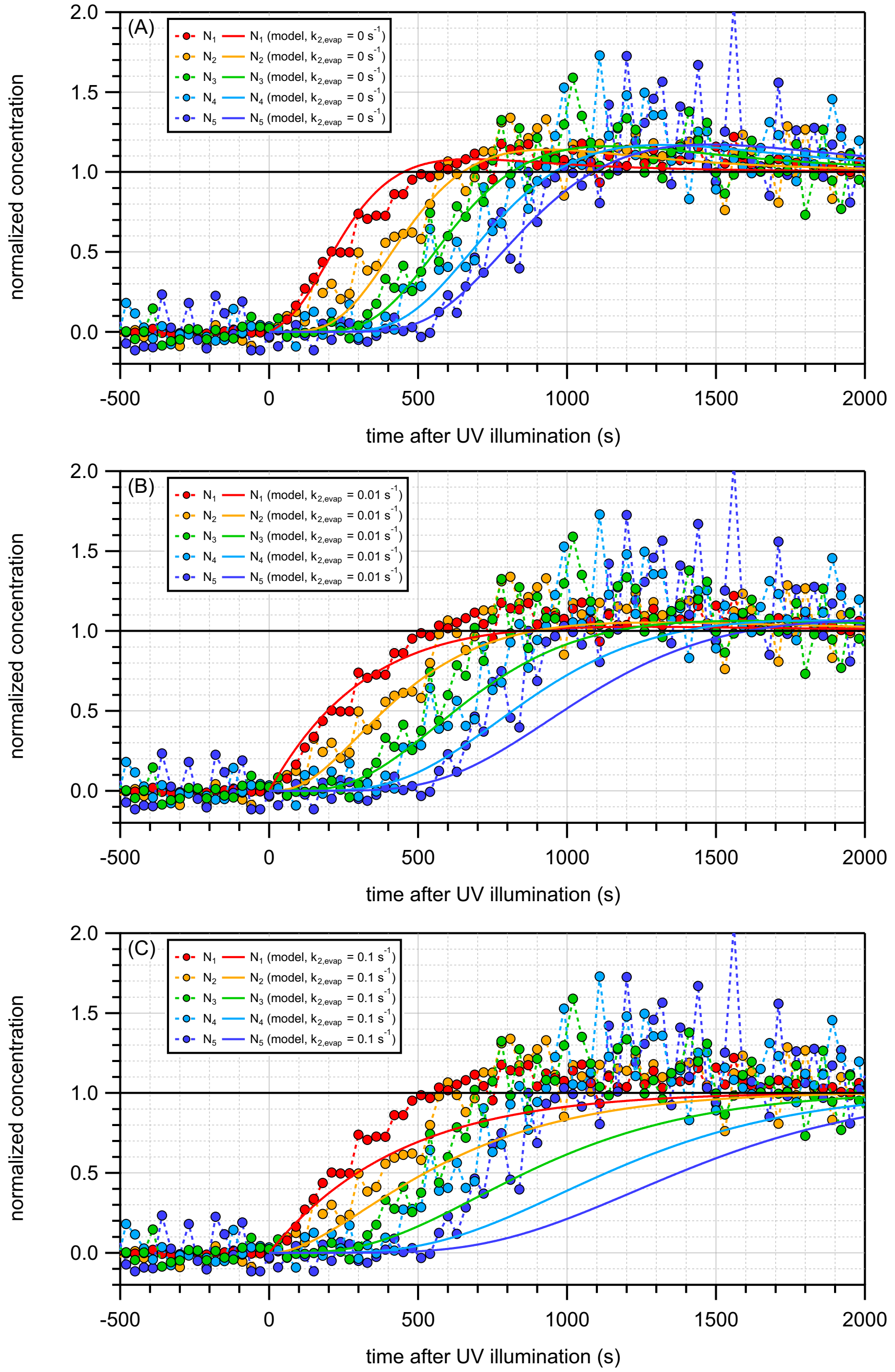

**Fig. S4.** Normalized cluster concentrations ($N_1$ to $N_5$) measured by the CI-APi-TOF ($N_1 = 2.2 \times 10^6$ cm$^{-3}$). Model calculations are shown by the solid lines assuming different dimer evaporation rates [$k_{2,evap} = 0$ s$^{-1}$ (*A*), $k_{2,evap} = 0.01$ s$^{-1}$ (*B*), and $k_{2,evap} = 0.1$ s$^{-1}$ (*C*)].